\documentclass[unsortedaddress,prd]{revtex4-1}

\usepackage{amsmath}
\usepackage{amsfonts}
\usepackage{amsthm}
\usepackage{amssymb}
\usepackage{graphicx}
\usepackage{hyperref}
\usepackage{color}

	\newcommand{\ncd}{\newcommand}
	\ncd{\mrm}    {\mathrm}
	\ncd{\beq} {\begin{equation}}
	\ncd{\eeq} {\end{equation}}

	\def\d{{\rm d}}

\begin{document}

	\title{Representation invariant Geometrothermodynamics: applications to ordinary thermodynamic systems}
	
\author{Hernando Quevedo}
	\email{quevedo@nucleares.unam.mx}
	\affiliation{Instituto de Ciencias Nucleares\\
     			 Universidad Nacional Aut\'onoma de M\'exico, A. P. 70-543, M\'exico D. F. 04510, M\'exico \\
     			 Instituto de Cosmologia, Relatividade e Astrofisica ICRA -CBPF \\
     			 Rua Dr. Xavier Sigaud, 150, CEP 22290-180, Rio de Janeiro, Brazil}

\author{Francisco Nettel}
	\email{fnettel@ciencias.unam.mx}
	\affiliation{Departamento de F\'\i sica, Fac. de Ciencias \\
			 Universidad Nacional Aut\'onoma de M\'exico, A. P. 50-542, M\'exico  D. F. 04510, M\'exico}

	\author{Cesar S. Lopez-Monsalvo}
	\email{cesar.slm@correo.nucleares.unam.mx}
	\affiliation{Instituto de Ciencias Nucleares\\
     			 Universidad Nacional Aut\'onoma de M\'exico, A. P. 70-543, M\'exico D. F. 04510, M\'exico}
	
	\author{Alessandro Bravetti}
		\email{bravetti@icranet.org}
		\affiliation{Dipartimento di Fisica and ICRA, "Sapienza" Universit\`a di Roma,\\ P.le Aldo Moro 5, I-00185 Rome, Italy}

	\date{\today}

	\begin{abstract}
	In this work we employ a recently devised metric within the Geometrothermodynamics program to study ordinary thermodynamic systems. The new feature of this metric is that, in addition to Legendre symmetry, it exhibits invariance under a change of representation. This metric was derived in a previous work by the authors while addressing the problem of the conformal structure of the thermodynamic metrics for different representations. Here, we present a thorough analysis for the ideal gas, the van der Waals fluid, the one dimensional Ising model and some other systems of cosmological interest. 
	\end{abstract}

\maketitle

\section{Introduction}

The description of thermodynamic systems by means of geometric objects associated to metrics has been examined for a long time. Thermodynamic metrics based on different physical grounds are widely used to describe phase transitions through curvature singularities. 
The first approach comes from information theory and in the thermodynamic limit leads to the Fisher-Rao metric \cite{rao45} which can be expressed as the Hessian of the entropy. Ruppeiner's metric, which originates from thermodynamic fluctuation theory and is expressed as (minus) the Hessian of the entropy \cite{ruppeiner}, is one of the most used metrics to analyze a wide range of thermodynamic systems.  It is known that this metric is not invariant under a change of representation. Indeed changing to the energy representation $U=U(E'^a)$, one obtains a metric which is conformal to the Weinhold metric given by the Hessian of the energy \cite{weinhold}, being the conformal factor the inverse of the temperature. Moreover, these metrics are  not invariant if one wishes to work in different thermodynamic potentials, i.e. the Massieu potentials or the various free energies, respectively.

A more recent attempt is that of the Geometrothermodynamics (GTD) program \cite{quevedo1}. The cornerstone of GTD is the invariance of its metrics under Legendre transformations. Thus, the GTD  description promotes the Legendre transformations to isometries. The metrics of the GTD program have been extensively applied to diverse physical systems, ranging from ordinary systems as the ideal gas and the van der Waals fluid \cite{quevedo2,tonio} to black hole thermodynamics \cite{quevedo3}. Recently, the issue of the invariance under a change of representation was addressed in a work on the conformal structure of the GTD metrics \cite{conformal}. The present work explores the geometric structure of various thermodynamic systems from a completely invariant point of view, i.e. we present a Legendre and representation invariant analysis.

In this paper we seek to establish the scope of this invariant metric under change of representation through its application to some physical systems. It is important to highlight the role of homogeneity when the change of representation is considered. Two representations are equivalent only if, at least one of the fundamental equations is a homogeneous function. Therefore, the hypothesis of homogeneity is essential in the construction of this metric.  

The rest of the article is organized as follows: in the section \ref{sec:gtd} we give a brief account of the GTD program and how we deal with the change of representation with the appropriate metric. Section \ref{sec:examples} is devoted to some applications of the invariant metric. We calculate the curvature scalar for the ideal gas, the van der Waals fluid and the one-dimensional Ising model and verify that the results correctly describe the thermodynamic interaction and the phase transitions, if any. At the end, in section \ref{sec:discussion}, we discuss the results and address some questions that remain to be answered.

\section{Representation and Legendre isometries in Geometrothermodynamics } 
\label{sec:gtd}

The underlying idea in the Geometrothermodynamics program is to construct a Legendre invariant Riemannian geometric theory to describe the thermodynamic properties of physical systems, e. g. phase transitions. On the one hand, it is conjectured that the curvature associated with the metric of the space of equilibrium states is related to some thermodynamic interaction, i.e. vanishing curvature would correspond to a non-interacting system, while a non-zero curvature signalizes the interaction between the components of the system. On the other hand, curvature singularities give account of the set of points where a phase transition occurs. 

Basically, there are two kind of metrics in the GTD program: those which are invariant under a total Legendre transformation (TLT) and the ones invariant under any partial Legendre transformation (PLT). Within the first group, there is a sub-classification of metrics which describe phase transitions of the first and second order separately.  Up to recently, none of these metrics was analyzed from the perspective of a change of representation. In a previous work by the authors \cite{conformal}, it was shown that only metrics invariant under TLT which are usually associated with first order phase transitions can be designed to give an invariant description in any representation, resulting in what is called the natural metric. Let us emphasize that two representations are thermodynamically equivalent if at least one of the possible fundamental relations is a homogeneous function of any order of the extensive variables $E^a$, $a=1,\ldots,n$. That is, if we denote the fundamental relation by $\Phi = \Phi(E^a)$, then 
	\beq
	\label{gtd.hom}
	\Phi(\lambda E^a) = \lambda^\beta \Phi(E^a), \quad  \text{where} \quad \lambda \in \mathbb{R}.
	\eeq

Trying to remain close to the purpose of the paper, we refer the reader interested in the analysis of the conformal structure of the GTD metrics and the proof of the invariance of the natural metric under a change of representation to the previous work \cite{conformal}. Here, we only present the invariant metric under total Legendre transformations and change of representation. A brief account of GTD is pertinent and so we proceed.

From a geometric point of view, Legendre transformations form a subset of the class preserving the contact structure of a $(2n+1)$-dimensional space called the \emph{thermodynamic phase space}, which we denote by the pair $(\mathcal{T},\Theta)$. Here $n$ represents the number of degrees of freedom (the number of extensive variables) and $\Theta$ plays the role of the canonical 1-form defining the contact structure $\xi \subset T\mathcal{T}$, that is, a maximally non-integrable family of hyperplanes satisfying \cite{mrugala1}
	\beq
	\label{gtd.01}
	\xi = \ker(\Theta), \quad \text{where} \quad \Theta \wedge (\d \Theta)^n \neq 0.
	\eeq
In practice, one adjusts the local coordinates of the thermodynamic phase space to be those for which  $\Theta$ takes the form
	\beq
	\Theta = \d \Phi - I_a \d E^a,
	\eeq
where $(\Phi, E^a,I_a)$, the coordinate functions of $\mathcal{T}$, correspond to the set of all the extensive and intensive variables together with the thermodynamic potential. Of special interest is the maximal integral sub-manifold, $\mathcal{E} \subset \mathcal{T}$, i.e. the largest sub-manifold which can be embedded in $\mathcal{T}$ such that $T\mathcal{E} \subset \xi$ \cite{mrugala2}. Alternatively, one can think of $\mathcal{E}$ as the embedded space given by the map
\beq  \label{embedding}
\varphi: \mathcal{E} \to \mathcal{T},
\eeq
where the isotropic condition $\varphi^*(\Theta) = 0$ must be met. It is easy to see that this is an $n$-dimensional manifold defined as the set for which the coordinate functions of $\mathcal{T}$ satisfy the condition
	\beq   \label{first}
	\d \Phi - I_a \d E^a = 0,
	\eeq
and as a consequence 
	\beq    \label{eos}
	I_a = \frac{\partial \Phi}{\partial E^a}.
	\eeq
It is now clear that $\Phi(E^a)$ is the fundamental relation for the thermodynamic system, $E^a$ the extensive variables, $I_a$ their corresponding intensive parameters and equations \eqref{first} and \eqref{eos} are the first law of thermodynamics and the set of equations of state, respectively. 

The above construction sets a clear footing in demanding Legendre symmetry. Legendre transformations are simply diffeomorphisms of $\mathcal{T}$ leaving the contact structure $\xi$, and therefore the space of equilibrium states $\mathcal{E}$, unchanged. Thus, if we endow the thermodynamic phase space with a Riemannian metric $G$ whose isometries are Legendre transformations, the curvature of the induced metric in the space of equilibrium states will be truly independent of the thermodynamic potential used to describe the system. There is, in addition, another symmetry implicit in the definition of the contact structure \eqref{gtd.01}. Any 1-form which is a scaling of the original $\Theta$ defines the same contact structure. Change or representations are particular cases of those scalings. 

It is possible to endow the contact manifold with a Legendre invariant metric $(\mathcal{T}, G)$, such that the pullback of $G$ induced by the embedding yields the thermodynamic metric in the space of equilibrium states 
\beq  \label{pullback}
\ g = \varphi^*(G).
\eeq

Going no further in the description of the formalism as it is done in \cite{conformal} we present at once the metric  which is invariant under Total Legendre Transformations and the change of representation, that is, 
	\beq  \label{Gnat}
	G^{\natural} = \Theta \otimes \Theta +\sum_{j \neq i} \frac{1}{E^j I_j} \d E^a \otimes \d I_a,
	\eeq
where we have excluded the $i$th pair of coordinates which is to be exchanged for the thermodynamic potential $\Phi$ when changing from one representation to the other.
We will call canonical representation to the thermodynamic fundamental relation which is a homogeneous function (of any order) in the extensive thermodynamic variables  [c.f. equation \eqref{gtd.hom}], $\Phi = \Phi(E^{(i)}, E^j)$, with $j= 1, \ldots,i-1, i+1, \ldots, n$, where $E^{(i)}$ is the extensive variable considered for an alternative or inverse representation $E^{(i)} = E^{(i)}(\Phi, E^j)$. Usually, these are the entropy and the internal energy, and for homogeneous systems of first order both are canonical representations. 

The induced metric on the space of equilibrium states $\mathcal{E}$ is
\beq   \label{gnat}
g^\natural  = \sum_{j \neq i} \left( E^j \frac{\partial \Phi}{\partial E^j} \right)^{-1} \frac{\partial^2 \Phi}{\partial E^b \partial E^a}\, \d E^a \otimes \d E^b.
\eeq
It should be clear from the above expression how to obtain the induced metric for the alternative representation. In the next section we will describe a few ordinary thermodynamic systems with the aid of the metric $g^\natural$ (natural metric). We will show that it is indeed invariant under a change of representation and that it correctly describes the expected thermodynamic behavior.

\section{Applications of natural metric to ordinary thermodynamic systems}
\label{sec:examples}

In this section we will revise some of the classical thermodynamic systems taking into account the invariance under change of representation. This aspect of the analysis has never been presented elsewhere and represents a step forward towards obtaining a fully invariant description of thermodynamics under the symmetries of the contact structure.

\subsection{The ideal gas}
\label{subsec:ig}

Let us  begin with the simplest example, the ideal gas. In this case, the fundamental relation in the entropy representation is given by \cite{callen}
\beq  \label{igs}
s(u, v) =  \left[\frac{3}{2}\ln u + \ln v \right],
\eeq
where we use molar quantities,  $R = \kappa_B N_A = 1$ with $\kappa_B$ the Boltzmann constant, $N_A$ the Avogadro number and $u$ and $v$ represent the internal molar energy and the molar volume of the system, respectively. The natural metric, equation \eqref{gnat}, 	for this representation reads
\beq   \label{igmets}
g^\natural_{s} = -\frac{3}{2}\frac{1}{u^2} \d u \otimes \d u - \frac{1}{v^2} \d v \otimes \d v,
\eeq
and its associated scalar curvature vanishes, in agreement with the curvature/thermodynamic interaction hypothesis. 

Working in the energy representation we have
\beq \label{igu}
u(s,v) = \left( \frac{1}{v} e^s \right)^{\frac{2}{3}},
\eeq
and the corresponding metric takes the form
\beq  \label{igmetu}
g^\natural_{u} = -\frac{2}{3} \d s \otimes \d s - \frac{5}{3 v^2} \d v \otimes \d v + \frac{4}{3 v} \d s \otimes \d v.
\eeq
As expected its Ricci curvature scalar also vanishes. Using any of the fundamental relations, equations \eqref{igs} or \eqref{igu}, it is straightforward to verify that the metrics \eqref{igmetu} and \eqref{igmets} are the same geometric object. This can also be done for the Gibbs free energy $g(T,P) = u - Ts + Pv$ as it comes from the total Legendre transformation applied to $u$. In view of the above agreement, it is an interesting exercise to explore the geometric behaviour if one attempts to work instead with a thermodynamic potential obtained from  a partial Legendre transformation. For instance, in the Helmholtz free energy $F(T,v)$ representation 
	\beq  \label{Fpot}
	F(T,v) = u - T s = \frac{1}{2}  T \left[3 - 2 \ln v - 3 \ln \frac{3 }{2} T \right],
	\eeq
the metric \eqref{gnat} takes the explicit form
	\beq   \label{gF}
	g^\natural_F (T,v) = \frac{3}{2 T^2} \d T \otimes \d T - \frac{1}{v^2} \d v \otimes \d v + \frac{2}{T v} \d T \otimes \d v.
	\eeq
If we write the expression above  in terms of the coordinates $\{s, v\}$ it becomes
\beq   \label{gFSV}
g^\natural_F (s,v) = \frac{2}{3} \d s \otimes \d s - \frac{5}{3 v^2} \d v \otimes \d v.
\eeq
Not surprisingly, it yields a different metric to \eqref{igmetu}, as $G^\natural$ is not invariant under partial Legendre transformations. Nevertheless, an interesting feature is that the scalar curvature for this potential vanishes as well. It is easy to verify that for the enthalpy $H(s,P) = u + P v$ the same situation is repeated. We will work out more complicated cases in which  this feature is no longer reproduced, e.g. the van der Waals fluid which presents a first order phase transition. In this case we will see that the geometric description is exactly the same for both representations $s$ and $u$,  and correctly signalizes the phase transition as a curvature singularity, while for a potential related to $u$ through a partial Legendre transformation (e.g. $F(T,v)$) we have that, being the natural metric a different object, the geometric description of the phase transition might not be correct.  

\subsection{The van der Waals system}
\label{subsec:vdw}

The fundamental equation for the van der Waals fluid in the entropy representation is given by
\beq   \label{vdw}
s(u,v) = \frac{3}{2} \ln  \left(u + \frac{a}{v} \right) + \ln \left( v -b \right).
\eeq
The metric is given by
	\begin{align}  \label{vdwmet}
	g^\natural_s = \frac{3 v^2 (v -b)}{(uv +a )\left[2uv^2 - a(v -3b) \right]} \d u \otimes \d u & + \frac{6 a (v -b)}{(uv +a ) \left[ uv^2 - a(v -3b) \right]} \d u \otimes \d v \nonumber\\ 
																								 & + \frac{2u^2 v^4 - (v^2 - 6bv +3b^2)(2a uv + a^2)}{v^2(v-b)(uv+a)(2uv^2 - av + 3ab)} \d v \otimes \d v,
	\end{align}
and the scalar curvature obtained is
	\beq  \label{vdwR}
	R^\natural_s =\frac{N^{\rm vdW}_s(u,v)}{4 \left(3 a b-a v+2 u v^2\right) \left(a \left(-3 b^2+6 b v-2 v^2\right)+u v^3\right)^2}
	\eeq
where the numerator function is 
	\begin{align}
	N^{\rm vdW}_s(u,v) = & \left[ a^3 \left( 27 b^5-243 b^4 v+504 b^3 v^2 -378 b^2 v^3 +113 b v^4-11 v^5\right)  \right. \nonumber\\ 
						 &  + 2 a^2 u v^2 \left(-72 b^4+174 b^3 v-111 b^2 v^2+16 b v^3+v^4\right) \nonumber \\ 
						 & \left. +4 a u^2 v^4 \left(-3 b^3+12 b^2 v-11 b v^2+v^3\right)-8 b u^3 v^7\right].
	\end{align}

 In the energy representation we have
	\beq \label{vdwu}
	u(s,v) = \frac{a(v-b)^{\frac{2}{3}} - v e^{\frac{2}{3}s}}{v(v-b)^{\frac{2}{3}}}.
	\eeq
In this case, the metric takes the form
	\begin{align}  \label{vdwumet}
	g^\natural_u = \frac{4 e^{2s/3} v (v-b)}{9a(v-b)^{5/3} -6 e^{2s/3} v^2} \d s \otimes \d s & + \frac{8 e^{2s/3} v}{6 e^{2s/3} v^2 - 9a(v-b)^{5/3}} \d s \otimes \d v \nonumber\\
																							  & + \frac{2 ( 5 e^{2s/3} v^3  - 9 a (v-b)^{8/3})}{3v^2 (v-b) ( 2 e^{2s/3} v^2 - 3a (v-b)^{5/3})} \d v \otimes \d v.
	\end{align}
and  the curvature is given by
	\beq  \label{vdwRu}
	R^\natural_u = \frac{N^{\rm vdW}_u(s,v)}{4 (v-b)^{1/3} \left(2 e^{2 s/3} v^2-3 a (v-b)^{5/3}\right) \left(e^{2 s/3} v^3-3 a (v-b)^{8/3}\right)^2} 
	\eeq
where the numerator is now 
	\begin{align}
	N^{\rm vdW}_u(s,v) = & \left[ -9 a^3 (v-b)^{16/3} \left(3 b^2-2 b v+v^2\right) \right.\nonumber  \\
						 & - 6 a^2 e^{2 s/3} v^2 (v-b)^{11/3} \left(24 b^2-14 b v+v^2\right) \nonumber  \\ 
						 & \left. +4 a e^{4 s/3} v^4 \left(3 b^4-15 b^3 v+17 b^2 v^2-6 b v^3+v^4\right)-8 b e^{2 s} v^7 (v-b)^{1/3} \right].
	\end{align}

It is a straightforward calculation to substitute the fundamental relation, equation \eqref{vdw}, in the above equation to obtain the same scalar curvature we get form the entropy representation, equation \eqref{vdwR}.  

In order to analyze the first order phase transition of the van der Waals system we start from \eqref{vdwR}. Using the equations of state
	\beq
	\frac{1}{T} = \frac{\partial s}{\partial u} \quad \text{and} \quad \frac{P}{T} = \frac{\partial s}{\partial v}
	\eeq
we obtain
\beq \label{eosp}
P = \frac{2 u v^2 - a v + 3 ab}{3 v^2 (v-b)},
\eeq 
which we can use to express the curvature scalar in the entropy representation, equation \eqref{vdwR}, as a function of the pressure and the volume, that is,
\begin{multline}  \label{vdwRvP}
R^\natural (v, P) = \frac{1}{3 P v^2 (v-b) \left(2 a b-a v+P v^3\right)^2} \bigg[-a^2 P v^2 \left(18 b^3-5 b^2 v-4 b v^2+v^3\right) \\
 - a^3 (v-6b) (v-2 b)^2 - a P^2 v^4 \left(-3 b^3+21 b^2 v-14 b v^2+v^3\right)+3 b P^3 v^7 (v-b) \bigg].
\end{multline}

It is a well known fact that the van der Waals system presents a first phase transition at points where \cite{callen}
\beq \label{vdwphase}
2 ab - a v + Pv^3 = 0.
\eeq
It is immediate to see that the scalar curvature obtained from the natural metric \eqref{vdwRvP} diverges at the same critical values; the denominator clearly vanishes at these points, while the numerator, 
\beq  \label{numR}
\text{Num}\left[R^\natural (v_c, P_c)\right] = -\frac{1}{v_c^2} \left[ a^3 (v_c-2 b)^2 \left(-9 b^3+21 b^2 v_c-13 b v_c^2+v_c^3\right)\right],
\eeq
remains finite.

It is worth mentioning that the additional points where the denominator  of the curvature scalar \eqref{vdwRvP} vanishes lack of physical meaning, as one can observe from the fundamental equation \eqref{vdw}. 

Now we show explicitly how $G^\natural$ fails to provide us a correct description of the thermodynamic behaviour of the van der Waals system if one attempts to use it starting with the Helmholtz free energy $F(T,v)$. This is easily understood by  noting that $F(T,v)$ cannot be considered the canonical representation nor its related (inverse) representation, given that it depends on $T$ which is not an extensive parameter evading the notion of homogeneity. The Helmholtz free energy is
\beq \label{free}
F(T,v) = \frac{1}{2 v} \left[ 3\left(Tv - \ln \frac{3 T}{2} \right) - 2(a + Tv \ln (v -b)\right].
\eeq 
In this case, the natural metric is
	\begin{align} \label{freemet}
	g^\natural_F =  \frac{3v(v-b)}{2T\left[a(v-b) - Tv^2 \right]}\ \d T \otimes \d T &+ \frac{2 v}{Tv^2 - a(v-b)}\ \d T \otimes \d v \nonumber\\
	               & + \frac{T V^3 - 2a(v-b)^2}{v^2 (v-b) \left[a(v-b) - Tv^2\right]}\ \d v \otimes \d v,
	\end{align}
and its associated curvature scalar is written as
	\beq
	\label{RnatF}
	R^\natural_F (T,v) = \frac{ N_F^\natural(T,v)}{\left[Tv^2 - a(v-b)\right]\,\left[6a(v-b)^2 - 5Tv^3\right]^2},
	\eeq
where
	\begin{align}
	 N_F^\natural(T,v) = & \left[ -15b T^3 v^7 - 3 a^2 T (v-b)^2 v^2 \left(v^2 - 14 b v + 24 b^2\right)\right. \nonumber\\
						 &\left. - 3a^3 (v-b)^3 \left(v^2 - 2bv + 3b^2\right) + aT^2 v^4 \left( 5v^3 - 25 b v^2 + 54 b^2 v - 9 b^3\right) \right].
	\end{align}
Using the equations of state one can express this curvature in terms of the pressure and the volume to obtain
\begin{multline}
R^\natural_F (v,P) = -\frac{1}{Pv^2 (v-b) (5Pv^3 - av + 6ab)^2} \bigg[a^3(v-2b)(v-6b)^2 - 15b P^3 v^7 (v-b) \\ 
- a P^2 v^4 (5v^3 - 70bv^2 + 99b^2 v - 9b^3) - aPv^2 (7v^3 - 50bv^2 + 39b^2 v + 54 b^3) \bigg].
\end{multline}
Evidently, this curvature scalar does not coincide with the one obtained from the canonical representations, nor its description of the phase transition is accurate. This is a consequence of the non-invariant property of the natural metric under partial Legendre transformations.

	\begin{figure}
	\includegraphics[width=0.7\columnwidth ]{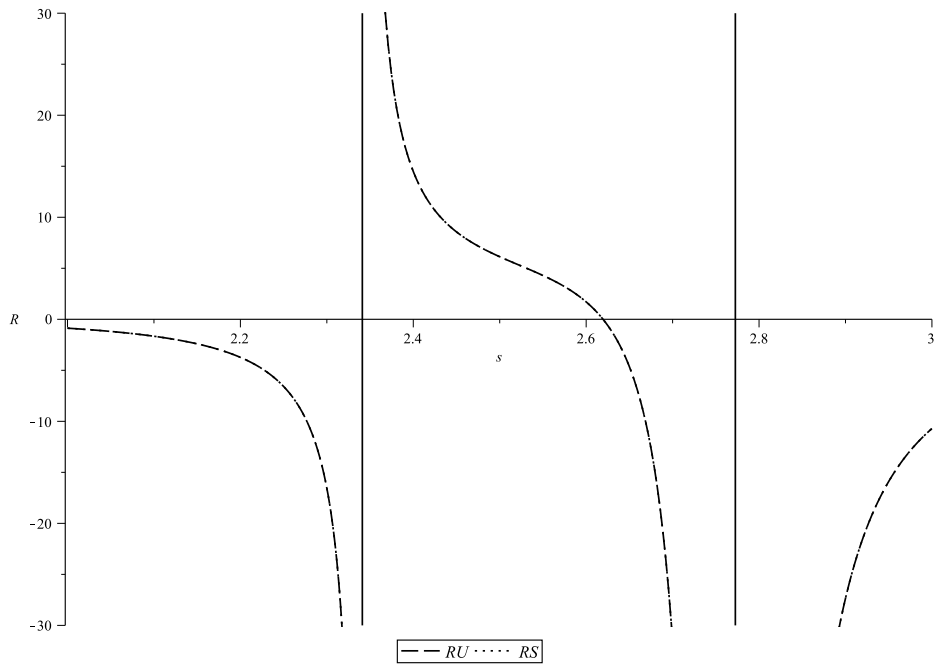}
	\caption{This figure shows the curvature scalar for the van der Waals system in both, energy and entropy representations. As expected, they are superimposed and show divergences at the phase-transition points. Here, we are using reduced variables.}
	\label{vdW1}
	\end{figure}
	\begin{figure}
	\includegraphics[width=0.7\columnwidth ]{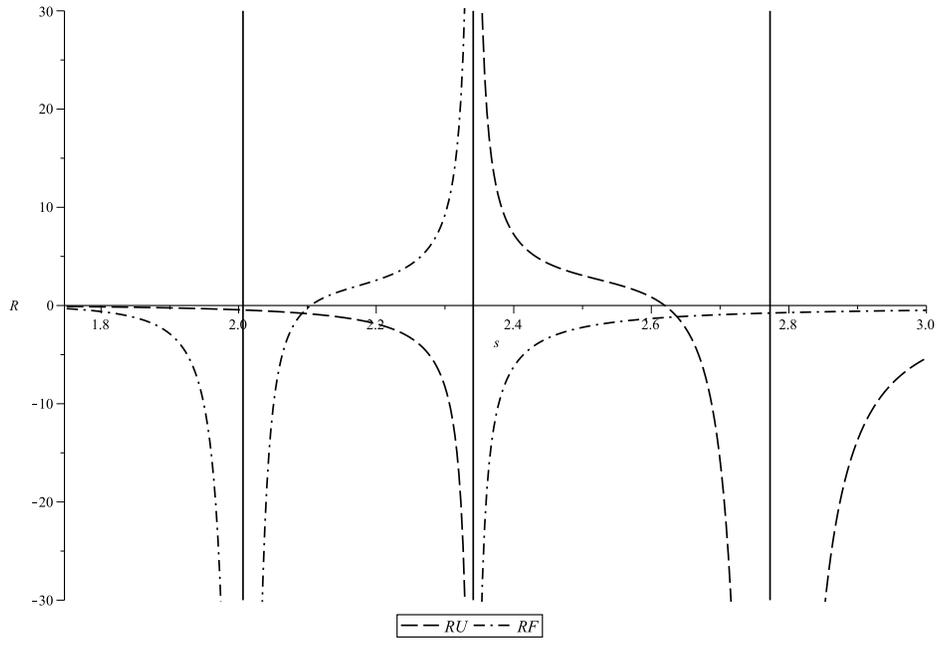}
	\caption{This figure shows the curvature scalar for the van der Waals system in both, energy and Helmholtz free-energy, representations. Here we observe that the curvatures do not coincide. Moreover, the scalar curvature $R^\natural_F$ does not diverge at the phase-transition point.}
	\label{vdW2}
	\end{figure}

\subsection{The one-dimensional Ising model}
\label{subsec:ising}
A slightly more complicated example is that of the one-dimensional Ising model. It is well known that there are no phase transitions for this system, albeit the statistical origin of its fundamental relation presents non-trivial thermodynamic interaction. In this case, the free energy obtained from the partition function is written as
	\beq
	\label{fising}
	f(\beta,H) =  - T \ln\left[\cosh\left(\frac{H}{T}\right) + \sqrt{\sinh^2\left(\frac{H}{T} \right)  + \exp\left(-\frac{4J}{T} \right)} \right],
	\eeq
where $H$ is the magnetic field, $\beta$ is the inverse of the temperature $T$ and $J$ measures the intensity of the interaction between the spins. Note that both, $T$ and $H$ are \emph{intensive} parameters, thus the free-energy \eqref{fising} corresponds to the total Legendre transformation of the internal energy $U$. Therefore, using the invariance of the natural metric under such a transformation, we can freely use this potential instead of the internal energy.   
	
In this case, the scalar curvature has the form
	\beq
	\label{Rnatising}
	R^\natural = \frac{1}{HT}\left[ \cosh\left(\frac{H}{T}\right) + \sqrt{\cosh^2\left(\frac{H}{T} \right) - 1 + \exp\left(-\frac{4 J }{T} \right)}\right]^{-3} \left[\sinh\left(\frac{H}{T}\right) \exp\left(\frac{8 J}{T}  \right) \right]^{-1} \ N(T,H),
	\eeq
where $N(T,H)$ corresponds to the numerator of $R^\natural$ whose explicit expression is not very illuminating for our argument. 

\begin{figure}
	\includegraphics[width=0.7\columnwidth ]{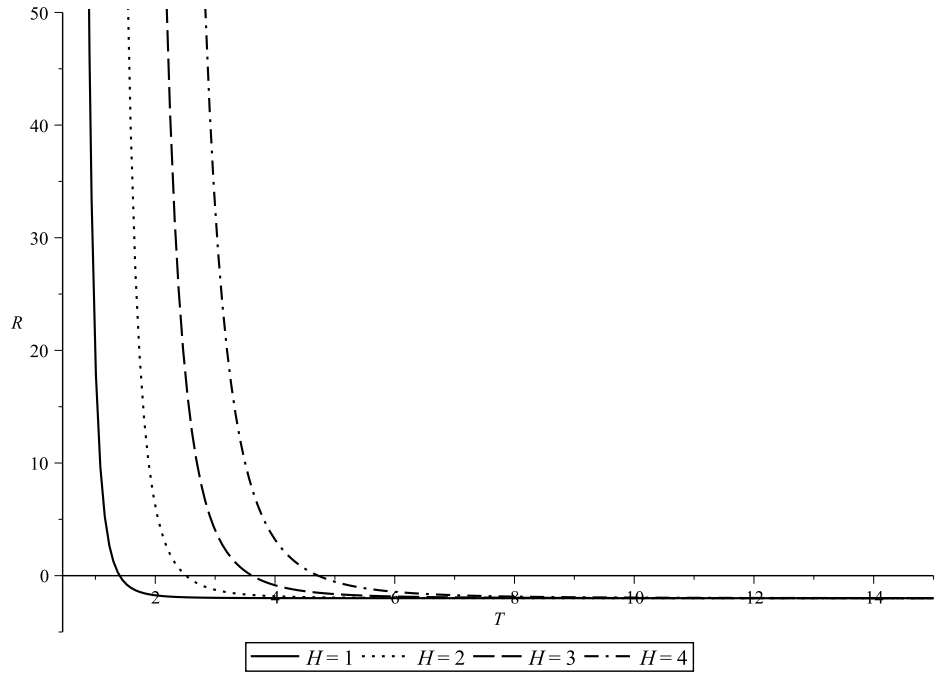}
	\caption{This figure shows various values of the curvature scalar for distinct intensities of the magnetic field. The one-dimensional Ising model does not present a curvature singularity other than at $T=0$, which is excluded by the thrird Law of thermodynamics. For large $T$ the curvature becomes asymptotically constant.}
	\label{figising}
	\end{figure}

According to our analysis, should we find that $R^\natural$ becomes singular this would indicate a phase-transition. However, it is easy to show that this only happens in the limit when $T$ goes to zero. In figure \ref{figising} we show a numeric analysis of the curvature scalar \eqref{Rnatising} where we observe that, indeed, the only divergence occurs at $T=0$. Moreover, the curvature becomes asymptotically constant for larger values of $T$. Thus, as we knew, the one-dimensional Ising model corresponds to an interacting system from the thermodynamic point of view, but without phase transitions. 

 \subsection{Generalized Chaplygin gas and the dark fluid equation}
\label{subsec:chap}	

Recently, within the realm of cosmological applications of GTD \cite{aztlan}, it has been proposed a fundamental relation depending on a pair of parameters which encloses the cases of the generalized Chaplygin gas and a dark fluid which can mimic the phenomenology of the  $\Lambda$CDM model. The fundamental equation in the entropy representation is
	\beq \label{chap}
	s(u,v) = s_0 \left( \ln u^{1 + \alpha} + C \ln v^{1 + \beta} \right),
	\eeq
where $\alpha$ and $\beta$ are real constants that determine the type of fluid we are dealing with. 

In this case, the expression for the metric is not very illuminating. However, it is interesting to note that  the expression for the metric density,
	\beq
	\label{densg}
	\det(g^\natural_S) =  \frac{\left[C \left(\alpha-\beta\right) v^{1+\beta} + \beta e^s\right] e^s}{C(1+\alpha) (1+\beta) v^{3+\beta} \left(-e^s + C v^{\alpha + \beta} \right) },
	\eeq
identically vanishes in the case $\alpha = \beta = 0$ and thus the metric is degenerate. Interestingly, this corresponds exactly to the $\Lambda$CDM model.  

The curvature scalar in the entropy representation is given by
	\beq \label{chapR}
	R^\natural_s = -\frac{(\beta +1)^2 \left(C^2 \alpha  v^{2 \beta +2}+2 C \alpha  u^{\alpha +1} v^{\beta +1}+\beta  u^{2 \alpha +2}\right)}{2 \left(C \alpha  v^{\beta +1}+\beta  u^{\alpha +1}\right)^2},
	\eeq
whereas in the energy representation one obtains
	\beq \label{chapRu}
	R^\natural_u = -\frac{(\beta +1)^2 \left(-2 C e^{\frac{s}{s_0}} v^{\beta +1} (\beta -\alpha )+\beta  e^{\frac{2 s}{s_0}}+C^2 v^{2 \beta +2} (\beta -\alpha )\right)}{2 \left(\beta e^{\frac{s}{s_0}}-C v^{\beta +1} (\beta -\alpha )\right)^2}.
	\eeq
It is an easy task to verify that the curvature is the same in both representations. 

An interesting feature of this  is that, in the case when $\alpha = \beta$, the curvature scalar is constant, 
	\beq \label{RchapAigualB}
	R^\natural_{\alpha =\beta} = - \frac{1}{2}\frac{(1 + \alpha)^2}{\alpha}.
	\eeq

Spaces of constant curvature are important and interesting in physics. A full analysis of them lies beyond the present work. However, it is worth mentioning that the case of constant curvature with $0 \leq \alpha \leq 1$ reproduces the equations of state for the generalized Chaplygin gas, while $\alpha = \beta = 1$ corresponds to the dubbed variable Chaplygin model. Moreover, when $\alpha = \beta \leq 0$  one obtains generic equations of state for various polytropic fluids. 

\section{Conclusions}
\label{sec:discussion}

The issue of representation invariance for metric theories of thermodynamics has not been previously addressed. In this paper we have worked out some applications of a recently derived metric within the GTD program which is invariant under a change of fundamental representation [c.f. equations \eqref{Gnat} and \eqref{gnat}]. As shown in a previous work  \cite{conformal}, in order to use the formalism correctly, the fundamental relation must be a homogeneous function of a definite order. This, of course, depends on the representation unless we consider only homogeneous functions of order one. This is a crucial point of the whole argument and it should be taken into account in any application of the metric \eqref{Gnat}. In all cases, one has to start the analysis in the representation where \eqref{gtd.hom} is satisfied or, alternatively, in its total Legendre transformed potential. 

We have shown that the metric \eqref{gnat} correctly accounts for the thermodynamic interaction and phase transition structure of a number of physical systems. In particular, we analyzed the thermodynamic geometry of the ideal gas, the van der Waals fluid, the one-dimensional Ising model and a fundamental relation with recent application in cosmology describing the generalized Chaplygin gas or a dark fluid \cite{aztlan} from a completely invariant point of view.

The case of the van der Waals gas allowed us to show what happens  if one starts from a fundamental relation which is not written in the canonical representation. As expected, the metric fails to give the correct thermodynamic description of the system. We calculated the thermodynamic curvature from the Helmholtz free energy $F=F(T,v)$, which is the result of applying a partial Legendre transformation to the canonical representation $u=u(s,v)$. In this case, the singularities of the curvature signal the occurrence of a phase transition which is in disagreement with the known results \cite{callen}, whereas working in the canonical representation we recover the correct phase-transition structure. This is clearly understood if one recalls how the natural metric was constructed.

The situation with the one-dimensional Ising model was a bit more subtle. The fundamental relation \eqref{fising} is a function of the magnetic field intensity and the temperature, both intensive variables. Here we used the fact that $G^\natural$ is invariant under total Legendre transformations to infer that the curvature associated with the metric in such a coordinate system would correctly account for the thermodynamic interaction as if an expression for the internal energy were available. 

Finally, the case of the generalized Chaplygin gas opened the possibility of analyzing  the class of thermodynamic potentials  producing a space of equilibrium states of constant curvature. Such analysis will be carried out elsewhere.

In sum, we have presented a few applications of a metric which, in addition to the Legendre symmetry, also exhibits representation invariance. These illustrate the correct use of the formalism, highlighting its subtleties and limits of applicability. This has lead us to classify the set of metrics in the GTD program according to the pursued objective when studying a specific thermodynamic system. Thus, if the interest is that of working in every possible potential available given a fixed thermodynamic representation, one is bound to use the metric which is invariant under partial Legendre transformations (c.f. $G_{\rm P}$ in \cite{conformal}). For applications to black-hole thermodynamics it remains an open matter the applicability of a variant of the invariant metrics under total Legendre transformations alone \cite{quevedo3}. Finally, the metric presented in this work  gives us the freedom to change thermodynamic representation and make total Legendre transformations as long as the system is homogeneous. 

\section*{Acknowledgements}

This work was supported by CONACYT, Grant No. 166391 and DGAPA-UNAM, Grant No. IN106110. FN receives support from DGAPA-UNAM (postdoctoral fellowship).  CSLM is thankful to CONACYT, postdoctoral Grant No. 290679\_UNAM. AB was supported by an ICRANet fellowship.

\bibliography{gnatreferences}

\begin{thebibliography}{99}


\bibitem{rao45} C. R. Rao, {\it Information and the accuracy attainable in the estimation of statistical parameters},
Bull. Calcutta Math. Soc. {\bf 37}, 81 (1945).

\bibitem{ruppeiner}
G. Ruppeiner,
{\it Thermodynamics: A Riemannian geometric model}, Phys. Rev. A {\bf 20}, 1608 (1979).

\bibitem{weinhold}
F. Weinhold,
{\it Metric geometry of equilibrium thermodynamics}, J. Chem. Phys. {\bf 63}, 2479 (1975).


\bibitem{quevedo1}
H. Quevedo,  
{\it   Geometrothermodynamics}, J. Math. Phys. {\bf 48}, 13506 (2007).


\bibitem{quevedo2}
H. Quevedo and M. N. Quevedo, {\it Fundamentals of Geometrothermodynamics}, Electronic J. Theor. Phys., Zacatecas Proceedings, pp. 1-16 (2011); 
arXiv:1111.5056 [gr-qc].


\bibitem{tonio}
H. Quevedo and A. Ramirez,
{\it A geometric approach to the thermodynamics of the van der Waals system}, (2012); arXiv:1205.3544 [math-ph].

\bibitem{quevedo3}
J. L. Alvarez, H. Quevedo and A. Sanchez,
{\it Unified geometric description of black hole thermodynamics}, Phys. Rev. D {\bf 77}, 084004 (2008).

\bibitem{conformal}
A. Bravetti, C. S. Lopez-Monsalvo, F. Nettel and H. Quevedo,
{\it The conformal metric structure of geometrothermodynamics}, J. Math. Phys. {\bf 54}, In press (2013); arXiv:1302.6928 [math-ph].

\bibitem{mrugala1}
R. Mrugala, {\it Geometrical formulation of equilibrium phenomenological thermodynamics}, Rep.
Math. Phys. {\bf 14}, 419 (1978).


\bibitem{mrugala2}
R. Mrugala, {\it Submanifolds in the thermodynamic phase space}, Rep. Math. Phys. {\bf 21}, 197 (1985).


\bibitem{callen}
H. B. Callen, {\it   Thermodynamics and an Introduction to Thermostatics}
(John Wiley and Sons, Inc., New York, 1985).


\bibitem{aztlan}
A. Aviles, A. Basterrechea-Almodovar, L. Campuzano and H. Quevedo,
{\it Extending the generalized Chaplygin gas model by using geometrothermodynamics}, Phys. Rev. D {\bf 86}, 063508 (2012).







\end{thebibliography}

\end{document}